\documentclass[floatfix,onecolumn,aps,prl,showpacs,amsmath,amssymb]{revtex4-1}
\usepackage[latin1]{inputenc}
\usepackage[dvips]{graphicx}

\usepackage{dcolumn}
\usepackage{bm}
\usepackage{dsfont}
\usepackage{color}
\usepackage{url}
\usepackage[colorlinks=true ,citecolor=blue]{hyperref}
\usepackage{amsmath,amssymb,esint}
\usepackage[table]{xcolor}

\definecolor{Nathanblue}{rgb}{0.,0.24,0.51}

\newcommand{\be}{\begin{equation}}
\newcommand{\ee}{\end{equation}}
\newcommand{\bq}{\begin{eqnarray}}
\newcommand{\eq}{\end{eqnarray}}

\begin{document}

\title{Emergent Dynamics of Spacetime and Matter from a Topological Phase}

\author{Giandomenico Palumbo}
\affiliation{School of Theoretical Physics, Dublin Institute for Advanced Studies, 10 Burlington Road,
	Dublin 4, Ireland}

\date{\today}

\email{giandomenico.palumbo@gmail.com}

\begin{abstract}
\noindent In this paper, we present a new theoretical scenario in which both dynamical Dirac fermions and Einstein's gravity with a positive cosmological constant and torsion emerge via a spontaneous symmetry breaking in a topological phase. This phase does not contain any local propagating degrees of freedom and is described by a metric-independent fermionic gauge theory, which is invariant under the de Sitter group. After breaking this group to its Lorentz subgroup through a dynamical Higgs mechanism, we show that fermions become dynamical and massive while the dynamics of the locally Lorentz-invariant spacetime can be induced via the condensation of the spinor field. Within our theory, both the Dirac mass and the Newton's constant of gravitation naturally emerge through well-known physical mechanisms.
\end{abstract}

\maketitle

\noindent {\bf{Introduction}}.
\noindent There has always been an intense debate about the origin of mass and the nature of gravity since Galileo and Newton. A deeper understanding of gravity as a fundamental force and of the mass as an intrinsic feature of fundamental matter would allow us to have a glimpse about the unifying theory of gravity and quantum field theory. \\ Although general relativity has passed a large number of experimental tests, it remains a classical theory and cannot explain the fundamental origin of the positive cosmological constant that is responsible of the accelerated expansion of the Universe.
At the same time, the Dirac theory of fermions and the Higgs mechanism, which are at the core of the Standard Model of particles, cannot explain the truly nature of neutrino masses \cite{Peskin}. \\
Well-recognized physical paradigms, such as the \emph{spontaneous symmetry breaking}, \emph{fermion condensation} and \emph{topological phases} play a crucial role not only in high-energy physics but also in condensed matter systems \cite{Volovik}. 
Here, we will employ these general concepts to build a different theoretical scenario for dynamical spacetime and matter that does not involve any quantum gravity theory \cite{Kiefer}. \\
Geometry and dynamics are tightly interconnected due to the central role of the metric tensor in propagating matter and gravity. Moreover, spinors are deeply related to the geometry of the underlying physical space while differential-geometric theories rely on the existence of topological spaces \cite{Nakahara}. \\
Although topological phases in gravity have been already discussed in literature \cite{Capovilla, Starodubtsev1, Starodubtsev2,Palumbo1,Palumbo2,Palumbo3, Gu1,Gu2,Gukov,Riotto,Volovik2}, their role in a more general picture that embraces both gravity and fermions is still lacking. \\
The main aim of this paper is to try to fill this gap by showing how propagating geometric and fermionic degrees of freedom can be consistently induced starting from a suitable topological phase.
Among the several implications of our theory, we will show that the Dirac mass of fermions and the Newton's constant of gravitation can be recognized as emergent physical quantities.\\

\noindent {\bf{Topological phase and gauge theory}}.
We start introducing a general family of fermionic gauge theories that are described by the following action
\begin{eqnarray}\label{spinor-BF2}
	S_{\Psi\Omega}\left(\overline{\Psi},\Psi, \Omega\right)=\int d^{4}x\,
	\overline{\Psi}\left( \overleftarrow{D}_\mu \Omega^{\mu\nu} \overrightarrow{D}_{\nu}\right) \Psi,
\end{eqnarray}
where $\mu,\nu=\{0,1,2,3\}$, $\Omega^{\mu\nu}$ is a differential operator, $\Psi_{\mu}$ is a fermionic field and $\overline{\Psi}_{\mu}$ its conjugate, $\overrightarrow{D}_{\mu} \Psi=i\partial_{\mu}\Psi+A_{\mu} \Psi$,  $\overline{\Psi} \overleftarrow{D}_\mu= i\partial_{\mu}\overline{\Psi}+ \overline{\Psi} A_{\mu}$, with $A_{\mu}$ the gauge connection that takes values in some Lie algebra $\mathcal{G}$ and
the integration is done on a 3+1-dimensional manifold $M_{4}$. The arrows over the symbol of the covariant derivative $D_\mu$ specify to which fermionic field they apply.
The gauge invariance of the above action is naturally satisfied if the fermionic fields and the differential operator $\Omega^{\mu\nu}$ transform as follows
\begin{eqnarray}
	\Psi_{\mu}\rightarrow U({\bf{x}})\, \Psi_\mu, \hspace{0.3cm} \overline{\Psi}_\mu\, \rightarrow \overline{\Psi}_\mu\, U({\bf{x}})^{\dagger}, \hspace{0.3cm} \Omega^{\mu\nu}\rightarrow U({\bf{x}})\Omega^{\mu\nu}U({\bf{x}})^{\dagger},
\end{eqnarray}
where $U(\bf{x})$ belongs to $\mathcal{G}$. A possible choice for $\Omega^{\mu\nu}$ is given by
\begin{eqnarray}
	\Omega^{\mu\nu}\equiv \, \epsilon^{\mu\nu \lambda \sigma}  (\overleftarrow{D}_\lambda  \Phi \overrightarrow{D}_{\sigma}+F_{\lambda\sigma}),
\end{eqnarray}
where $\epsilon^{\mu\nu\alpha\beta}$ is the Levi-Civita symbol, $\Phi$ is a $\mathcal{G}$-valued complex scalar field, while 
\begin{eqnarray}
	F_{\mu\nu} = \partial_{\mu}A_\nu-\partial_{\nu}A_\mu  + i [ A_\mu, A_\nu],
\end{eqnarray}
is the curvature tensor of the gauge connection. Under gauge transformations, both $F_{\mu\nu}$ and $\Phi$ transform in a covariant way.
We can then rewrite $S_{\Psi\Omega}$ as follows
\begin{eqnarray}\label{spinor-BF3}
	S_{\Psi TFT}\left(\overline{\Psi}, \Psi, A, \Phi \right)= \int d^{4}x\, \epsilon^{\mu\nu \lambda \sigma}\, \overline{\Psi}\left(F_{\mu\nu} \Phi F_{\lambda\sigma}+ \overleftarrow{D}_\mu F_{\nu\lambda} \overrightarrow{D}_{\sigma}\right) \Psi,
\end{eqnarray}
where we have used $\overline{\Psi} \left[\overleftarrow{D}_\mu, \overleftarrow{D}_\nu \right]= \overline{\Psi} F_{\mu\nu}$ and $\left[\overrightarrow{D}_\mu, \overrightarrow{D}_\nu \right] \Psi = F_{\mu\nu}\Psi$.
The action is metric independent and describes a topological phase.
In other words, the fermion field $\Psi$ has no local propagating degrees of freedom because 
of the absence of a metric tensor $g_{\mu\nu}$ in the action. Notice that a similar action without $\Phi$ was already introduced in Ref.\cite{Palumbo2}. Although $g_{\mu\nu}$ is a central quantity in the second-order formalism of general relativity, in this paper we will adopt the first-order formalism, in which the tetrads $e_{\mu}^{a}$ are the fundamental variables of spacetime, such that $g_{\mu\nu}=e_{\mu}^{a}\,e_{\nu}^{b}\,\eta_{ab}$, where $\eta_{ab}$ is the Minkowski flat metric and $a,b=\{0,1,2,3\}$ \cite{Blagojevic}. \\

\noindent {\bf{De Sitter symmetry breaking and dynamical fermions}}. 
We now identify $\mathcal{G}$ with $SO(4,1)$, i.e. the de Sitter group. Notice that this gauge group, which is compatible with the existence of a positive cosmological constant at global level, has been already employed to build some dynamical gravitational models and quantum field theories \cite{MM,Stelle,Wise,Randono, Randono2, Westman, Westman2, Fukuyama,Ikeda}.
The spinorial representation of de Sitter group is given in terms of $4 \times 4$ matrices $\Sigma_{AB}=\,i\,[\gamma_{A},\gamma_{B}]/4$, where $\gamma_A= (\gamma_a, i \gamma_5)$ are the Dirac matrices and $\gamma_{5}=i\gamma_{0}\gamma_{1}\gamma_{2}\gamma_{3}$ is the chiral matrix \cite{Randono,Westman}. 
By fixing the group, we can consequently define the expression of the conjugate field, i.e. $\overline{\Psi}= \Psi^\dagger \gamma_0$ \cite{Westman}.
The spontaneous symmetry breaking of the de Sitter group occurs through the dynamical Higgs mechanism, i.e. by considering an Higgs action term $S_H(\Phi)$ for the scalar field $\Phi \equiv \phi_A \gamma_A$ \cite{Stelle}, such that we can define a total action
\begin{eqnarray}\label{spinor-BF9}
	S_T= S_{\Psi TFT}\left(\overline{\Psi}, \Psi, A, \Phi \right) + S_H(\Phi).
\end{eqnarray}
In the Lorentz-preserving phase, $\Phi$ acquires the following non-zero expectation value \cite{Fukuyama,Ikeda}
\begin{eqnarray}
	\langle \Phi \rangle = -i \tilde{\phi}_5 \gamma_5,
\end{eqnarray}
with $\phi_a = 0$ and $\tilde{\phi}_5$ a real constant parameter. This expression preserves the invariance under the $SO(3,1)$ Lorentz subgroup. To keep the gauge invariance of the whole action under the Lorentz transformations, the fermion fields now have to transform as  $\Psi\rightarrow U_L({\bf{x}}) \Psi$,  $\overline{\Psi}\rightarrow \overline{\Psi} U_L(\bf{x})^{\dagger}$
where $U_L(\bf{x})$ belongs now to $SO(3,1)$. 
Moreover, $A_{\mu}$ is not only a gauge connection but also a Cartan connection \cite{Wise,Randono}. Cartan geometry represents a natural generalization of Riemannian geometry with non-zero torsion \cite{Blagojevic}.  Importantly, in Cartan geometry, there exists a natural decomposition of the Cartan connection that is given by \cite{Randono,Westman}
\begin{eqnarray}\label{decomposition}
	A_{\mu}=\omega_{\mu}+\frac{1}{2\,l}\,\gamma_{5}\widehat{\gamma}_{\mu},
\end{eqnarray}
where $l$ is a dimensionful constant parameter (i.e. the de Sitter length), $\omega_{\mu}=(1/2)\omega_{\mu}^{ab}\Sigma_{ab}$ with $\Sigma_{ab}=\,i\,[\gamma_{a},\gamma_{b}]/4$ is the Lorentz-invariant spin connection and $\widehat{\gamma}_{\mu}=e_{\mu}^{a}\gamma_{a}$, such that $\{\widehat{\gamma}_{\mu},\widehat{\gamma}_{\nu}\}=2 g_{\mu\nu}\mathbb{I}$. Here, differently from torsionless general relativity, $\omega_{\mu}^{ab}$  and $e_{\mu}^{a}$ are independent fields. In this way, the corresponding curvature tensor $F_{\mu\nu}$ is given by
\begin{eqnarray}\label{decomposition2}
	F_{\mu\nu}=R_{\mu\nu}
	-\frac{i}{4\,l^{2}}[\widehat{\gamma}_{\mu},\widehat{\gamma}_{\nu}]-\frac{1}{2\,l}\gamma_{5} T_{\mu\nu},
\end{eqnarray}
where $R_{\mu\nu}\equiv R_{\mu\nu}^{ab}\Sigma_{ab}$ and $T_{\mu\nu}\equiv  T^a_{\mu\nu}
\gamma_a$ are the Riemann and torsion tensors, respectively.
We can now collect all the terms in Eq.(\ref{spinor-BF3}) that contain the product of three and four $\widehat{\gamma}_\mu$'s, i.e.
\begin{align}\label{equations}
	\epsilon^{\mu\nu\lambda \sigma}\overline{\Psi}\left(-\frac{i}{4\,l^{3}}\gamma_{5}\widehat{\gamma}_{\mu}\widehat{\gamma}_{\nu}\widehat{\gamma}_{\lambda}(i \partial_{\sigma} +\omega_{\sigma})+\frac{i}{8\,l^{4}} \widehat{\gamma}_{\mu}\widehat{\gamma}_{\nu}\widehat{\gamma}_{\lambda}\widehat{\gamma}_{\sigma}+
	\frac{i \tilde{\phi}_5}{4 l^4}\gamma_5\widehat{\gamma}_{\mu}\widehat{\gamma}_{\nu}\widehat{\gamma}_{\lambda}\widehat{\gamma}_{\sigma}\right)\Psi,
\end{align}
and show that they give rise to metric-dependent terms. In fact, due to the following identity
\begin{eqnarray}
	\gamma_{a}\gamma_{b}\gamma_{c}=\eta_{ab}\gamma_{c}+\eta_{bc}\gamma_{a}- \eta_{ca}\gamma_{b}+i \,\epsilon_{abcd}\, \gamma_{5}\gamma^{d},
\end{eqnarray}
we can rewrite the totally anti-symmetric products of tetrads in Eq.(\ref{equations}) \cite{Palumbo2,Westman}
\begin{eqnarray}\label{product}
	\epsilon^{\mu\nu\lambda\sigma}\,\gamma_{5}
	\widehat{\gamma}_{\mu}\widehat{\gamma}_{\nu}
	\widehat{\gamma}_{\lambda}=
	- 3!\, i |e| \widehat{\gamma}^{\sigma}, \hspace{0.4cm} \epsilon^{\mu\nu\lambda\sigma}\,\gamma_{5}
	\widehat{\gamma}_{\mu}\widehat{\gamma}_{\nu} 
	\widehat{\gamma}_{\lambda} \widehat{\gamma}_{\sigma}=
	- 4!\, i |e|,
\end{eqnarray}
where $|e|=-(1/4!)\,\epsilon^{\mu\nu\alpha\beta}\epsilon_{abcd}\,e_{\mu}^{a}\,e_{\nu}^{b}\,e_{\alpha}^{c}\,e_{\beta}^{d}$ is the determinant of the tetrads, having $\epsilon^{\mu\nu\alpha\beta}\epsilon_{\mu\nu\alpha\beta}=-4!$ as convention.
By applying the above identities in Eq.(\ref{equations}), we can easily see that they give rise to the following action
\begin{eqnarray}\label{Dirac}
	S_{Dirac}\left(\overline{\psi},\psi, \omega_{\sigma}\right)=-\int d^{4}x\,|e|\,
	\overline{\psi}\,(i\,\widehat{\gamma}^{\sigma}\partial_{\sigma} +\widehat{\gamma}^{\sigma} \omega_{\sigma} -\gamma_{5}m_{C}- m_D)\psi+ h.c. + ...
\end{eqnarray}
which is nothing but the Dirac action for the (dimensioful) spinor fields $\overline{\psi} \equiv (\sqrt{3/2 l^3})\overline{\Psi}$ and $\psi \equiv (\sqrt{3/2 l^3})\Psi$,  where $m_{C}=2/l$ is the chiral mass and
\begin{eqnarray}
	m_{D}=4 \tilde{\phi}_5/l
\end{eqnarray}
is the \emph{Dirac mass}. This last expression shows that relativistic fermions can acquire a Dirac mass even in absence of other fundamental interactions.
Notice that in Eq.(\ref{Dirac}), $...$ represent all the other terms in Eq.(\ref{spinor-BF3}) that depend on $R_{\mu\nu}$, $T_{\mu\nu}$ and $\omega_{\mu}$. Because these terms remain metric independent, they do not influence the fermionic dynamics and vanish in the flat spacetime limit ($\omega_{\mu}\rightarrow 0$). \\

\noindent {\bf{Fermion condensation and dynamical gravity}}.
We now reconsider the action in Eq.(\ref{spinor-BF3}), where $\Psi$ has been replaced by a different fermionic species $\chi$ that undergoes a fermion condensation \cite{Volovik,Randono2}.
After the de Sitter symmetry breaking, Eq.(\ref{spinor-BF3}) can be rewritten as follows
\begin{eqnarray}\label{spinor-BF5}
	S_{\chi}= \int d^{4}x\, \epsilon^{\mu\nu \lambda \sigma}\, \left[  {\rm Tr} \left( -i \tilde{\phi}_5 \gamma_5  F_{\mu\nu} F_{\lambda\sigma} \chi \bar{\chi}\right)+\bar{\chi} \overleftarrow{D}_\mu F_{\nu\lambda} \overrightarrow{D}_{\sigma} \chi \right],
\end{eqnarray}
where the trace is taken on the gauge index while $\bar{\chi}$ and $\chi$ transform under the Lorentz group. Similarly to fermion condensation analyzed in Ref.\cite{Randono2}, here
the spinor fields form a condensate when the matrix $\chi \bar{\chi}$
reduces to the vacuum expectation value 
\begin{eqnarray}
	\langle  \chi \bar{\chi} \rangle = \theta \mathbb{I}, 
\end{eqnarray}
where $\theta$ is a real constant parameter and $\mathbb{I}$ is the identity matrix. In this regime, the second term in $S_{\chi}$ vanishes. Due to Eq.(\ref{decomposition2}), the first term in Eq.(\ref{spinor-BF5}) can be equivalently rewritten as
\begin{eqnarray}\label{spinor-BF6}
	\epsilon^{\mu\nu \lambda \sigma}\,  {\rm Tr} \left( -i  \tilde{\phi}_5 \theta \gamma_5 F_{\mu\nu} F_{\lambda\sigma} \right)= \hspace{8.0cm} \nonumber \\
	-i  \tilde{\phi}_5 \theta\, \epsilon^{\mu\nu \lambda \sigma}\left[  \left(- \frac{1}{4}R^{ab}_{\mu\nu}R^{cd}_{\lambda\sigma}+\frac{1}{2 l^2} R^{ab}_{\mu\nu}e^c_\lambda e^d_\sigma -\frac{1}{4 l^4} e^a_\mu e^b_\nu e^c_\lambda e^d_\sigma\right) {\rm Tr}( \gamma_5 \gamma_a \gamma_b \gamma_c \gamma_d)+\right. \nonumber \\
	\left. \frac{1}{4 l^2} T^a_{\mu\nu}T^b_{\lambda\sigma}{\rm Tr}(\gamma_5 \gamma_a \gamma_b) 
	- \frac{i}{2 l} \left( R^{ab}_{\mu\nu}-\frac{1}{l^2} e^a_\mu e^b_\nu \right) T^c_{\lambda \sigma}
	{\rm Tr}( \gamma_a \gamma_b \gamma_c) \right].
\end{eqnarray}
By employing the following identities for the $4 \times 4$ Dirac matrices
\begin{eqnarray}
	{\rm Tr}( \gamma_5 \gamma_a \gamma_b \gamma_c \gamma_d) = -4 i \epsilon_{abcd}, \hspace{0.3cm}
	{\rm Tr}(\gamma_5 \gamma_a \gamma_b) =0, \hspace{0.3cm} {\rm Tr}( \gamma_a \gamma_b \gamma_c)=0,
\end{eqnarray}
we finally obtain
\begin{eqnarray}\label{spinor-BF7}
	S_{EH}=-\frac{1}{32 \pi G}\int d^{4}x\, \epsilon^{\mu\nu \lambda \sigma}\, \epsilon_{abcd}\left(-\frac{l^2}{2} R^{ab}_{\mu\nu}R^{cd}_{\lambda\sigma}+R^{ab}_{\mu\nu}e^c_\lambda e^d_\sigma -\frac{1}{2 l^2} e^a_\mu e^b_\nu e^c_\lambda e^d_\sigma\right),
\end{eqnarray}
where the first term is the topological Euler invariant while the second and the third terms are the Einstein-Hilbert and cosmological constant terms, respectively \cite{Blagojevic}.
Here, $\Lambda \sim 1/l^2$ is the positive cosmological constant that gives rise to the accelerated expansion of the Universe, while the \emph{Newton's constant}
\begin{eqnarray}
	G=l^2/ (64 \pi \tilde{\phi}_5 \theta)
\end{eqnarray}
is induced by the de Sitter length $l$, the symmetry-breaking parameter $\tilde{\phi}_5$ and the expectation value of the fermionic condensate $\theta$. This last expression clearly shows that even without invoking any dynamical theory of quantum gravity, some aspects
of classical gravity can have a quantum origin as already suggested by other emergent gravity theories \cite{Sakharov, Liberati, Verlinde,Padmanabhan}.


\begin{thebibliography}{99}
	
	\bibitem{Peskin}
	M. E. Peskin and D. V. Schroeder, \emph{An introduction to quantum field theory}, Addison-Wesley Press (1995).
	
	\bibitem{Volovik}
	G. E. Volovik, {\em The Universe in a Helium Droplet}, Oxford University Press (2003).
	
	\bibitem{Kiefer}
	C. Kiefer, \emph{Quantum Gravity}, Oxford University Press (2007).
	
	\bibitem{Nakahara}
	M. Nakahara, \emph{Geometry, Topology and Physics}, CRC Press (2003).
	
	\bibitem{Capovilla}
	R. Capovilla, M. Montesinos, V. A. Prieto and E. Rojas,
	Class. Quantum Grav. \textbf{18}, L49 (2001).
	
	\bibitem{Starodubtsev1}
	L. Smolin and A. Starodubtsev, arXiv:hep-th/0311163.
	
	\bibitem{Starodubtsev2}
	L. Freidel and A. Starodubtsev, arXiv:hep-th/0501191.
	
	
	\bibitem{Palumbo1}
	G. Palumbo, Mod. Phys. Lett. A \textbf{31}, 1650015 (2016).
	
	
	\bibitem{Palumbo2}
	G. Palumbo, EPL \textbf{114}, 50001 (2016).
	
	\bibitem{Palumbo3}
	G. Palumbo, European Physical Journal Plus \textbf{135}, 142 (2020).
	
	\bibitem{Gu1}
	Z.-C. Gu, arXiv:1709.09806.
	
	\bibitem{Gu2}
	T. Fang and Z.-C. Gu, arXiv:2106.10242.
	
	\bibitem{Gukov}
	P. Agrawal, S. Gukov, G. Obied and C. Vafa, arXiv:2009.10077.
	
	\bibitem{Riotto}
	A. Kehagias and A. Riotto, arXiv:2105.10669.
	
	\bibitem{Volovik2}
	F. R. Klinkhamer and G. E. Volovik, arXiv:2111.07962.
	
	
	\bibitem{Blagojevic}
	M. Blagojevic and F.W. Hehl, \emph{Gauge Theories and Gravitation}, Imperial College Press (2013).
	
	
	\bibitem{MM} 
	S. W. MacDowell and F. Mansouri, Phys. Rev. Lett. \textbf{38}, 739 (1977). Erratum, ibid. \textbf{38}, 1376 (1977).
	
	
	
	
	
	\bibitem{Stelle}
	K. S. Stelle and P. C. West, Phys. Rev. D \textbf{21}, 1466 (1980).
	
	
	\bibitem{Wise}
	D. K. Wise, Class. Quantum Grav. \textbf{27}, 155010 (2010).
	
	
	
	\bibitem{Randono}
	A. Randono, arXiv:1010.5822.
	
	
	
	\bibitem{Randono2}
	A. Randono, Class. Quantum Grav. \textbf{27}, 215019 (2010).
	
	
	
	\bibitem{Westman}
	H. F. Westman and T. G. Zlosnik, Ann. Phys. \textbf{334}, 157 (2013).
	
	\bibitem{Westman2}
	H. F. Westman and T. G. Zlosnik, Ann. Phys. \textbf{361}, 330 (2015).
	
	\bibitem{Fukuyama}
	T. Fukuyama, Ann. Phys. \textbf{157}, 321 (1984).
	
	\bibitem{Ikeda}
	N. Ikeda and T. Fukuyama, Prog. Theor. Phys. \textbf{122}, 339 (2009).
	
	\bibitem{Sakharov}
	A. D. Sakharov, Sov. Phys. Dokl. \textbf{12}, 1040 (1968).
	
	\bibitem{Liberati}
	C. Barcelo, M. Visser, and S. Liberati, Int. J. Mod. Phys. D \textbf{10}, 799 (2001).
	
	\bibitem{Verlinde}
	E. Verlinde, Journal of High Energy Physics \textbf{2011}, 29 (2011).
	
	\bibitem{Padmanabhan}
	T. Padmanabhan, Modern Phys. Lett. A \textbf{30}, 1540007 (2015).
	
\end{thebibliography}
\end{document}